\documentclass[prb,twocolumn,superscriptaffiliation,showpacs]{revtex4}

\usepackage{epsf,latexsym}
\usepackage{amssymb,amsmath}
\usepackage{times}

\newcommand{\ignore}[1]{}

\newcommand{\be}{\begin{equation}}
\newcommand{\ee}{\end{equation}}

\def\CC{{\rm\kern.24em \vrule width.04em height1.46ex depth-.07ex
    \kern-.30em C}}
\def\P{{\rm I\kern-.25em P}}
\def\RR{{\rm
         \vrule width.04em height1.58ex depth-.0ex
         \kern-.04em R}}

\def\bbbc{{\mathchoice {\setbox0=\hbox{$\displaystyle\rm C$}\hbox{\hbox
to0pt{\kern0.4\wd0\vrule height0.9\ht0\hss}\box0}}
{\setbox0=\hbox{$\textstyle\rm C$}\hbox{\hbox
to0pt{\kern0.4\wd0\vrule height0.9\ht0\hss}\box0}}
{\setbox0=\hbox{$\scriptstyle\rm C$}\hbox{\hbox
to0pt{\kern0.4\wd0\vrule height0.9\ht0\hss}\box0}}
{\setbox0=\hbox{$\scriptscriptstyle\rm C$}\hbox{\hbox
to0pt{\kern0.4\wd0\vrule height0.9\ht0\hss}\box0}}}}
\def\bbbz{{\mathchoice {\hbox{$\sf\textstyle Z\kern-0.4em Z$}}
{\hbox{$\sf\textstyle Z\kern-0.4em Z$}}
{\hbox{$\sf\scriptstyle Z\kern-0.3em Z$}}
{\hbox{$\sf\scriptscriptstyle Z\kern-0.2em Z$}}}}

\begin{document}

\title{Entanglement, fidelity and topological entropy in a quantum phase
transition to topological order}
\author{A. Hamma,$^{(1)}$ W. Zhang,$^{(2)}$ S. Haas,$^{(2)}$, and D.A. Lidar$%
^{(1,2,3)}$}
\affiliation{Departments of Chemistry,$^{(1)}$ Physics and
  Astronomy,$^{(2)}$ and Electrical Engineering,\!\!~$^{(3)}$\\
  Center for Quantum Information Science \& Technology\\
University of Southern California, Los Angeles, CA 90089, USA}

\begin{abstract}
We present a numerical study of a quantum phase transition from a
spin-polarized to a topologically ordered phase in a 
system of spin-$1/2$ particles on a torus. We demonstrate that this
non-symmetry-breaking topological quantum phase
transition (TOQPT) is of second order.
The transition is analyzed via the ground state energy and fidelity, 
block entanglement, Wilson loops, and the recently proposed
topological entropy. Only the topological entropy distinguishes the TOQPT from
a standard QPT, and remarkably, does so already for small system
sizes. Thus the topological entropy serves as a proper order parameter. We
demonstrate that our
conclusions are robust under the addition of random
perturbations, not only in the topological phase,
but also in the spin polarized phase and even at the critical point.
\end{abstract}

\maketitle

\section{Introduction}
A quantum phase transition (QPT) occurs when the order parameter of a 
quantum system becomes
discontinuous or singular \cite{Sachdev}. This is associated with a drastic change of the
ground state wave function. Unlike classical phase transitions, QPTs occur at 
$T=0$ and thus are not driven by thermal fluctuations. Instead, quantum
fluctuations are 
capable of changing the internal order of a system and cause the transition.
When a quantum Hamiltonian $H(\lambda )$, which depends smoothly on
external parameters $\lambda $, approaches a quantum critical point $%
\lambda _{c}$ from a gapped phase, the gap $\Delta $ above the ground state closes,
and the critical system has gapless excitations. This
corresponds to a continuous, second order QPT.

Here, we consider a QPT from a spin polarized to a topologically
ordered phase: a topological quantum phase transition (TOQPT).
The internal order that characterizes topologically ordered phases cannot be explained by the standard Ginzburg-Landau
theory of symmetry breaking and local order parameters. Instead, it
requires the notion of \emph{Topological Order} (TO) \cite{Wen}. 
TO manifests itself in a ground state degeneracy
which depends on the topology of the physical system, and it is robust
against arbitrary local perturbations \cite{Wen:90}. This robustness is at
the root of topological quantum computation, i.e., the ground state
degeneracy can be used as a robust memory, and the topological 
interactions
among the quasi-particles can be used to construct robust logic gates
\cite{Kitaev:03,Freedman:03}. On the other hand, to what extent a
TOQPT is affected by perturbations is a problem that has only very
recently been addressed \cite{HammaLidar:06,Trebst:06}, and is
a focus of this work.
Moreover, the classification of TO is still an open question. Ground state
degeneracy, quasiparticle statistics and edge states, all measure and detect
TO but do not suffice to give a full description.
Tools from quantum information theory, specifically entanglement \cite{entqpt,dft} and the ground state fidelity
\cite{zanardi}, have recently been
widely exploited to characterize QPTs. To date, all the QPTs studied
with these tools have been of the usual 
symmetry breaking type. Here we apply them to the transition from a
spin-polarized phase to a TO phase, and find that they are universal
in the sense that they detect this transition. 
However, these tools do not suffice to distinguish a
symmetry breaking QPT from a TOQPT.
Recently, the new concept of 
\textquotedblleft topological entropy\textquotedblright\
$S_{\mathrm{top}}$ was introduced 
\cite{Kitaev:06Levin:06}. The
topological entropy vanishes in the thermodynamic limit for a normal
state, whereas $S_{\mathrm{top}}\neq 0$ for a TO state. Therefore, 
$S_{\mathrm{%
top}}$ can serve as an order parameter. Moreover, TO is not
only a property of infinite systems, and an important question that was
left open in Refs.~\cite{Kitaev:06Levin:06} is the behavior of $S_{\mathrm{%
top}}$ for finite systems. Here we shed light on this question by presenting
finite-system calculations of
$S_{\mathrm{top}}$. We report that $S_{%
\mathrm{top}}$ changes abruptly at the critical point of a phase
transition between  phases with
and without TO, even for very small systems.
It is thus an excellent discriminator between the absence and presence of
TO, and moreover, $S_\mathrm{top}$ \emph{is capable of detecting a TOQPT}.

Specifically, we present an exact time-dependent numerical study of a TOQPT,
introduced in Ref.~\cite{HammaLidar:06}, from a spin-polarized  
phase to a TO\ phase, for both the ideal model and the model in presence of an external perturbation. 
Our results are the following: (i) standard QPT\
detectors (derivative of the ground state energy \cite{Sachdev},
entanglement of a subsystem with the remainder of the lattice 
\cite{entqpt,dft},
ground state fidelity \cite{zanardi}), are all singular at the critical
point of the TOQPT, thus confirming that
this is indeed a QPT. Ground state fidelity and block entanglement are
thus capable of dealing also with non symmetry breaking QPTs.  (ii) 
$S_{\mathrm{top}}$ detects the TOQPT 
in a very sharp manner already for small system sizes. It also detects TO
better than other non-local order parameters, in particular the expectation
value of Wilson loops. It is therefore appropriate for the detection and
characterization of TOQPTs and for studying TO. These results complement and
strengthen the conclusions of Ref.~\cite{Kitaev:06Levin:06}. (iii) 
Adiabatic evolution can initialize 
topological quantum memory faithfully: even in the presence of
perturbations the coupling to other 
topological sectors and excited states is negligible. (iv) This
robustness extends to the entire topological phase, and even to the
critical point itself. Perturbations do not affect the nature of the
TOQPT either.
\begin{figure}[tbp]
\begin{equation*}
\leavevmode\hbox{\epsfxsize=6 cm
   \epsffile{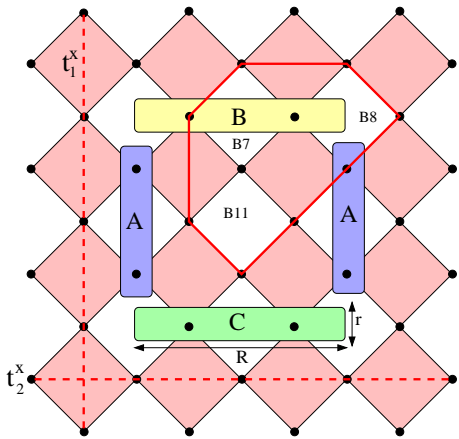}}
\end{equation*}%
\caption{(Color online)
A square lattice with $32$ spins. The 
spin degrees 
of freedom are placed on the
vertices. The red dashed lines $t_1^x, t_2^x$ are the incontractible
loops around the torus. The product $B7\cdot B8\cdot B11$ denotes the
loop operator  drawn in
red. All the spins on the vertices crossed by a loop are flipped. The region $%
A\cup B\cup C$ is a ring containing eight spins,
used in computing $S_{\rm top}$.
For the lattice of $32$ spins, the ring has diameter $R=2$ and width
$r=1$.}
\label{lattice}
\end{figure}

\section{Preliminaries}
Consider a square lattice $L$ with periodic boundary 
conditions (torus) and with $n$ spin-$1/2$
degrees of freedom occupying its vertices. The Hilbert space is given
by $\mathcal{H}=\mathrm{span}\{|0\rangle ,|1\rangle 
\}^{\otimes n}$, where $|0\rangle $ and $|1\rangle $ are the $\pm $
eigenvectors of the Pauli $\sigma ^{z}$ matrix. As shown in Fig. \ref%
{lattice}, the $n$ plaquettes can be partitioned into two sub-lattices, 
denoted by different colors. 
Following Kitaev \cite{Kitaev:03},
we associate with every
white plaquette $p$ an operator $B_{p}\equiv \prod_{j\in
\partial p}\sigma _{j}^{x}$ that flips all spins along the boundary of $p
$. A \textquotedblleft closed string operator\textquotedblright\ is a
product of plaquette operators $B_{p}$ that flips all spins around a
loop (or around a loop net). The \textquotedblleft group of closed
strings\textquotedblright\ $\overline{X}$ is the group of products of
plaquettes $B_{p}$. Similarly, with every pink plaquette $s$, we associate 
an 
operator $A_{s}\equiv \prod_{j\in s}\sigma _{j}^{z}$ which counts if
there is an even or odd number of flipped spins around the plaquette
$s$. Kitaev's toric code Hamiltonian \cite{Kitaev:03} is then given by  
$
H_{U,g}=-U\sum_{s}A_{s}-g\sum_{p}B_{p}\equiv H_{U}+H_{g},
$
which realizes a $Z_{2}$ lattice gauge theory in the limit $U\rightarrow
\infty $. The ground state is an equal superposition of all closed
strings (loops) acting on the spin polarized state $|\mathrm{vac}\rangle \equiv
|0\rangle _{1}\otimes ...\otimes |0\rangle _{n}$ -- it is in a \emph{%
string-condensed} phase. The ground state manifold is given by 
$
\mathcal{L}=\mbox{span}\{|\overline{X}|^{-\frac{1}{2}%
}(t_{1}^{x})^{i}(t_{2}^{x})^{j}\sum_{x\in \overline{X}}x|\mathrm{vac}\rangle
;\;i,j\in \{0,1\}\},
$
which is fourfold degenerate \cite{Hamma:05}. The 
$t_{1,2}^{x}$s flip all the spins along an incontractible loop around
the torus (See Fig.\ref{lattice}), taking a vector in 
$\mathcal{L}$ to an orthogonal one in the same manifold because they
commute with $H_{U,g}$. On a lattice on a Riemann surface of genus
$g$, there are $2g$ incontractible
loops $\{ t_{j}^{x}\}_{j=1}^{2g}$,
and therefore $\mathcal{L}$ 
is $2^{2g}$-fold degenerate
\cite{Kitaev:03,Hamma:05a} (for a torus $g=1$).

\textit{The Model and the QPT}.--- Now consider the following
time-dependent Hamiltonian,
introduced in \cite{HammaLidar:06} as a model for a TOQPT:
\begin{equation}
  H_0(\tau )=H_{U}+\tau H_{g}+(1-\tau )H_{\xi },
  \label{eq:htau}
\end{equation}%
where $H_{\xi }\equiv -\xi \sum_{r=1}^{n}\sigma _{r}^{z}$, $\tau
=t/T\in \lbrack 0,1]$, and $T$ is the total time. The non-degenerate
  ground state of $H(0)=H_{U}+H_{\xi}$ is the spin polarized state
  $|\mathrm{vac}\rangle $ which is the vacuum of the strings. 
The term $(1-\tau )H_{\xi }$ acts as a
tension for the strings, whereas $\tau H_{g}$ causes the strings 
to fluctuate. As $\tau $ increases, the string fluctuations increase while the loop tension
decreases. For a critical value of $\lambda \equiv \tau g/(1-\tau )\xi $, and in
the thermodynamic limit, a continuous QPT occurs to a TO phase of
string condensation. This QPT is not symmetry breaking, i.e., is a
TOQPT. As argued in Ref.~\cite{HammaLidar:06}, provided $T\gg 1/\Delta
_{\min }$ (the minimum gap, as a function of $\tau$, between the ground 
state
and the first excited state) evolution according to $H(\tau )$ is an
adiabatic preparation mechanism of a TO state:\ one of the $2^{2g}$
degenerate ground states of Kitaev's toric code model \cite{Kitaev:03}. Ref.~%
\cite{HammaLidar:06} showed that
$\Delta _{\min }\sim 1/\sqrt{n}$. $H(\tau )$ can be mapped onto an Ising model in a transverse field, which 
is 
known to have a second order QPT \cite{HammaLidar:06} (see also \cite{Trebst:06}). However, in this work we do not resort to such a
mapping, because it is non-local and does not preserve entanglement
measures. Instead, we
numerically study $H(\tau )$ for $\tau \in \lbrack 0,1]$ in $\Delta
  \tau =.01$ increments on lattices $Ln$ with $n=\{8,18,32\}$ spins,
and set $U=100,\xi = g =1$.
The computational methods used here are (i) the Housholder
algorithm \cite{Press} for the full diagonalization (all eigenstates) of $L8$, and (ii) a modified Lanczos method \cite{Gagliano} to obtain the
low-energy sectors of $L18$ and $L32$. We observe that for all $%
\tau \in \lbrack 0,1]$ the ground state comprises only closed strings.
Since this is the case for every finite system size, and in order to reduce
computation cost, we diagonalize $L32$ only in the relevant symmetry
subspaces, defined by the constraint $A_{s}|\psi \rangle \equiv \prod_{j\in
s}\sigma _{j}^{z}|\psi \rangle =|\psi \rangle $, $\forall s$. 

\section{The perturbed model}
To test the robustness of the TOQPT, we
also studied the perturbed model given by
\be
H(\tau) =H_0(\tau) + V\equiv H_0(\tau) + \sum_{j=1}^n \left( h^x(j)\sigma_j^x
+ h^z(j)\sigma_j^z \right)
\ee
The perturbation $V$ is random with $h^{z}(j)$ and $h^{x}(j)$
uniformly distributed in $[-0.2,0.2]$ and $[-P,P]$, respectively, with
the magnitude $P$ variable in our calculations below.
We carried out calculations for $L8$ (time-dependent)
and $L18$ (ground state only). These were
averaged over random realizations of $V$, and included the full
Hilbert space as $V$ disrupts the symmetry $A_{s}|\psi\rangle=|\psi
\rangle $. The $z$-component of the perturbation is expected to have a
small effect as it only slightly modifies the term $H_{\xi}$ for $\tau
< \tau_c$, while for $\tau > \tau_c$ TO dominates and tension effects
are suppressed. Our calculations confirmed this, and hence
Figs.~\ref{energyspectrum}-\ref{tope4} show the results for
$h^z(j)\equiv 0$.

\section{Adiabatic evolution}
We numerically simulated the time evolution from the fully  
polarized state at $\tau = 0$ to the string-condensed phase at $\tau =
1$. The possibility of preparation of topological order via such
  evolution has been studied theoretically in
  Ref.~\onlinecite{HammaLidar:06}. A crucial point is to show that the
  adiabatic time depends on the minimum gap that marks the phase
  transition (and that is polynomially small in the number of spins),
  and not on the exponentially small splitting of the ground state in
  the topological phase. To this end, one must show that transitions to other topological sectors are forbidden and protected by topology.\cite{HammaLidar:06} 
The initial wave function 
is the exactly known
ground state of $H(\tau =0)$. This
state is then used as the seed to compute the 
ground state of $H(\Delta \tau)$. After iteration, this 
state is in turn used as the seed for $H(2\Delta\tau)$, etc.
\begin{figure}[tbp]
\begin{equation*}
\leavevmode\hbox{\epsfxsize=8cm
    \epsffile{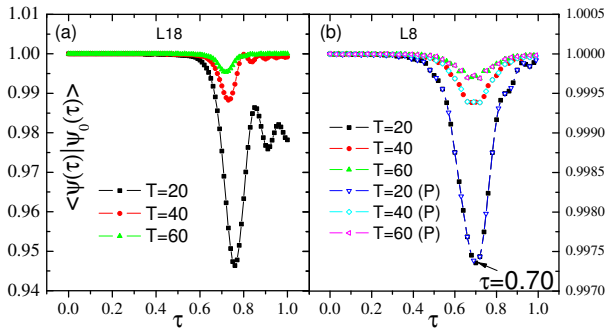}}
\end{equation*}
\caption{(Color online)
Fidelity between the
time-dependent solution of the Schr\"{o}dinger equation
and the adiabatic state,
for different values of the total evolution 
time: $T = 20, 40, 60$. (a) 
The unperturbed model for $L18$. The evolution is adiabatic for
$T=60$. Note that the drop in 
adiabaticity is a precursor of the QPT. (b)
$L8$: fidelity in
both the ideal and perturbed (P=1) cases.  The perturbed model
is indistinguishable from the ideal one.}
\label{energyspectrum}
\end{figure}
We can estimate
to what extent the evolution is adiabatic by numerically solving, for $L18$,  the 
time dependent Schr\"{o}dinger equation $ H\psi (\tau)= i \dot{\psi}(\tau)$
for different values of the total
evolution time $T$.
This is shown in Fig.~\ref{energyspectrum}(a), where
we plot the fidelity between the time
evolved wave function $\psi(\tau)$ and the instantaneous ground state: $\mathcal
F_{\rm ad} = |\langle\psi(\tau)|\psi_0(\tau)\rangle|$.
Moreover, we compute
${\cal F}_{\rm ad}$ also for the perturbed model, but
the largest lattice for which we can do this is $L8$.
Fig. 2b shows clearly that for $P=1$ the perturbation does not
change the time-evolved state. Significant effects start at $P=2$ (not shown).
We also find that the
overlap between the evolved wave function
$\psi(\tau)$ and the other sectors $%
(t_{1}^{x})^{i}(t_{2}^{x})^{j}|\mathrm{\psi_0(\tau)}\rangle $
is of order $\sim10^{-3}$ for every $(i,j)\ne (0,0)$
and value of $T$ tested.
This is numerical
evidence for the argument that time evolution will always keep the
instantaneous eigenstate within a topological sector, even in presence of perturbations
\cite{HammaLidar:06}. Thus the relevant gap for
adiabatic evolution is that to the other closed string excited states,
which implies that the evolution into the TO sector can be used to
prepare a topological quantum memory \cite{HammaLidar:06}.
Henceforth we work only in the sector $(i=0,j=0)$, into which the system
is initialized as the unique ground state of $H(\tau =0)$.

\section{Detecting the QPT with standard measures}
\begin{figure}[tbp]
\begin{equation*}
\leavevmode\hbox{\epsfxsize=9 cm
    \epsffile{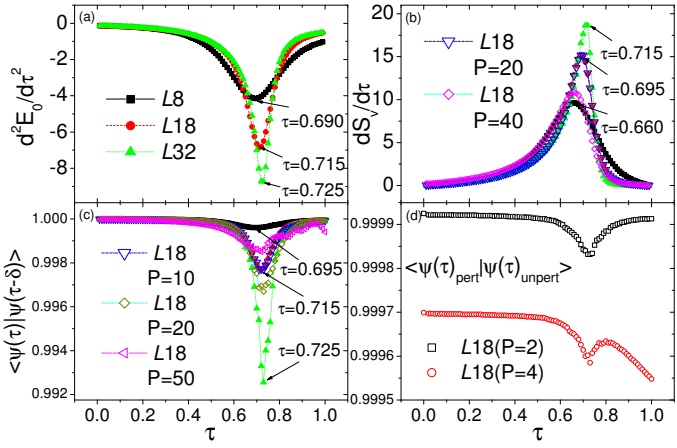}}
\end{equation*}%
\caption{(Color online)
QPT detectors for $L8,L18,L32$,
for the unperturbed and perturbed model. All graphs show strong
  resilience of the model and its QPT against perturbations: (a)
  Second derivative of $E(\protect\tau )$, diverging for $\protect\tau
  _{c}\sim .7$. The QPT is thus second order. (b) Derivative of the
  von Neumann entropy, measuring the entanglement of a plaquette with 
the rest of
the lattice. Its divergence at criticality also signals a
second order QPT.
The perturbation has no effect for $P=20$ (triangles indistinguishable
from circles) but is visible for $P=40$.
(c) Ground state
fidelity $\mathcal{F}(\protect\tau )$: the fidelity drop at the
critical point signals a QPT, associated with a drastic change in the
properties of the ground state. (d) Overlap between the perturbed and
the ideal ground state.
The clearly visible
susceptibility to the perturbation at the critical point also signals
  the QPT.
}
\label{state}
\end{figure}

To check that the
transition from magnetic order to TO is indeed a QPT, we first computed the
energy per particle of the ground state for $L8,L18,L32$, and its second
derivative. As seen in Fig.~\ref{state}(a), the latter develops a
singularity as system size increases, signaling a second order QPT with a
critical point at $\tau \sim 0.71$, corresponding to a ratio $\xi/g
\sim 0.41$.
This is in good agreement with the analytical study 
\cite{castelnovo}, which obtained (in the thermodynamic limit)
$\xi/g\sim .44$, even if this model is only asymptotically equivalent
to the toric code in a magnetic field, in the small field limit. On the
other hand, Ref.~\cite{Trebst:06} found $\xi/g \sim 0.33$, using 
a mapping to the classical 3D Ising model.
In Fig.~\ref{state}(b) we show the block entanglement between four spins
in a small loop ($B11$, Fig.~\ref{lattice}) and the rest of the lattice,
as measured by the von Neumann entropy. In agreement with the general
theory \cite{dft}, the derivative of the entanglement diverges at the
critical point for a second order QPT.

{A new interesting alternative characterization of QPTs can be
given in terms of the scaling in the fidelity $\mathcal{F}_{\Delta \tau }(\tau
)=|\langle \psi (\tau )|\psi (\tau -\Delta \tau )\rangle |$ between two
different ground states \cite{zanardi}. At a quantum phase transition, the fidelity should scale to zero superextensively. Previous work \cite{zanardi, campos} has shown that the fidelity criterion is valid for generic symmetry breaking second order QPTs. Nevertheless, the fidelity criterion is not strictly local, so one would like to know whether it detects the QPT to a topologically ordered state. The results are shown in Fig.~\ref{state}(c). The
fidelity drop criterion indeed also detects the QPT.
Figures~\ref{state}(a)-(c) also show the result for the perturbed
model.}

{By looking at the behavior of the transition in the presence of perturbations, we} can safely conclude that the QPT is unaffected by the perturbation
for $P\le 10$, 
namely the value of $\tau_c$ and the magnitude of the fidelity drop
remain unchanged.
In Fig.~\ref{state}(d),
we plot the overlap between the perturbed and unperturbed ground state.
The drop in this quantity also signals the
QPT, showing that the system is most sensitive to perturbations at the
critical point (see also Ref.~\cite{Quan:06}).
Interestingly, in contrast to the robustness of the entanglement and
$\mathcal{F}_{\Delta \tau }(\tau)$, the perturbed and unperturbed
ground states differ significantly already for $P>2$. 
The results in Fig.~\ref{state} thus allow
us to infer unambiguously that there is indeed a second order QPT in the
adiabatic dynamics generated by $H(\tau )$. However, none of the
quantities shown in Fig.~\ref{state} is explicitly designed to detect
topological features, and hence these
quantities are incapable of
distinguishing between a symmetry breaking
QPT and a TOQPT.

\begin{figure}[tbp1]
\begin{equation*}
\leavevmode\hbox{\epsfxsize=9 cm
    \epsffile{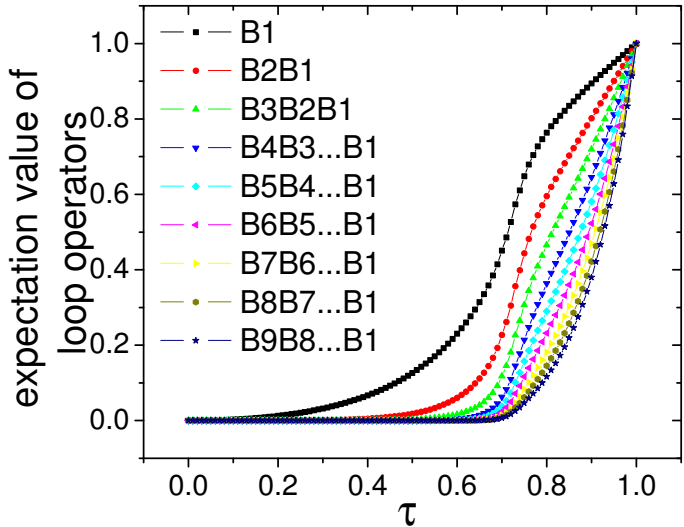}}
\end{equation*}%
\caption{(Color online) Expectation value of Wilson loop operators of
increasing size for $L32$
The expectation value of the loop operators starts to increase at
$\tau_c$, more steeply so for the largest loops, indicating that this
observable can be used to detect the TOQPT for large systems.}
\label{tope1}
\end{figure}
\begin{figure}[tbp2]
\begin{equation*}
\leavevmode\hbox{\epsfxsize=9 cm
    \epsffile{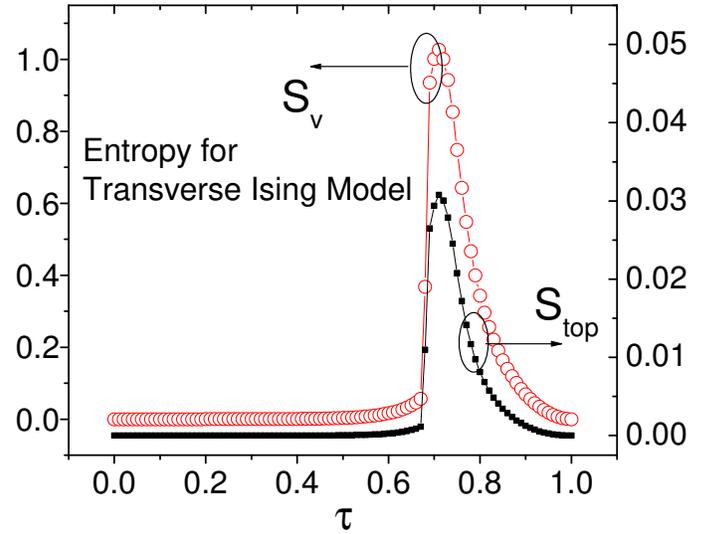}}
\end{equation*}%
\caption{(Color online) von Neumann entropy for a plaquette of spins 
and
$S_{\mathrm{top}}$ for
$L32$ with an Ising Hamiltonian in 
a transverse field. Note the different vertical scales. For a system
without topological order,
$S_{\mathrm{top}}$ is always $\sim 0$.
(the small bump is a finite-size effect).}
\label{tope2}
\end{figure}
\begin{figure}[tbp3]
\begin{equation*}
\leavevmode\hbox{\epsfxsize=9 cm
    \epsffile{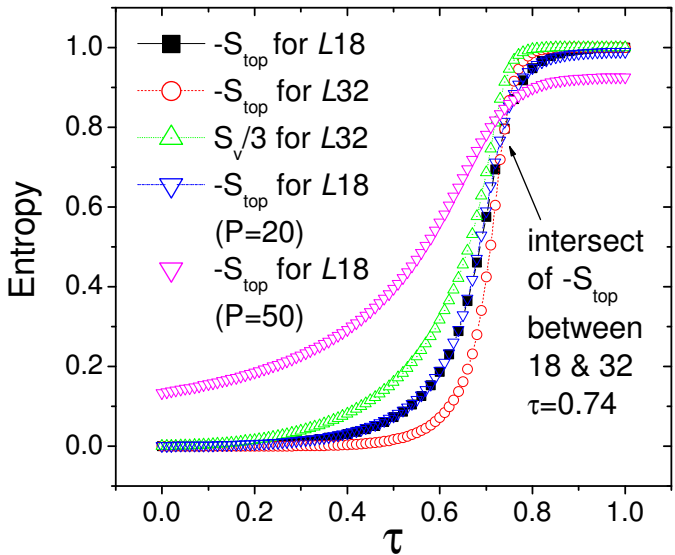}}
\end{equation*}%
\caption{(Color online) $S_{\mathrm{top}}$
for $L18$ and $L32$, 
and von Neumann entropy for
$L32$, for the ideal and perturbed model. $S_v$ assumes
the value $l-1=3$ in the entire TO phase, where $l-1$ is
the exact value of $S_v$ for the pure Kitaev model ($\tau=1$) and
$l=4$ is the length of the border of a plaquette.
$S_{\mathrm{top}}$ is 
zero in the spin-polarized phase and quickly reaches
unity in the
TO phase. Also the topological character of the QPT and of the phases is resilient to perturbations.}
\label{tope3}
\end{figure}
\begin{figure}[tbp4]
\begin{equation*}
\leavevmode\hbox{\epsfxsize=9 cm
    \epsffile{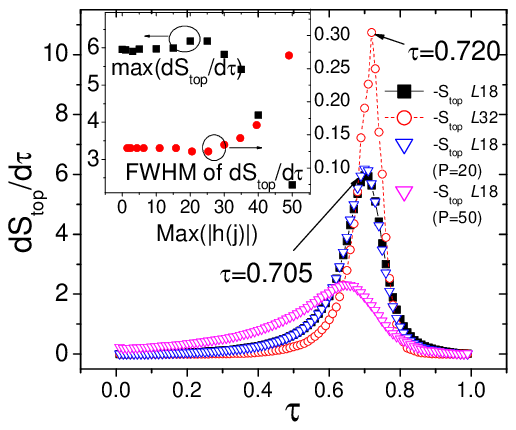}}
\end{equation*}%
\caption{(Color online) First
derivative of
$S_{\mathrm{top}}$ for
$L18$ and $L32$, for the ideal and perturbed model,
diverging at
$\tau_c$
thus signaling the TOQPT; (inset) derivative and full width at half
maximum of $S_{\rm top}$ at $\tau_c$
as function of
perturbation strength $P$. 
$S_{\rm top}$ remains robust up to $P \sim 25$. }
\label{tope4}
\end{figure}
\section{Characterizing the topological phase}
The 
spin-polarized regime for $\tau <\tau _{c}
$ is characterized by a finite magnetization. {On the other hand, the topologically ordered phase $\tau >\tau _{c}$ does not admit a local order parameter\cite{elitzur}. The topologically ordered phase is a string condensed phase and an effective $Z_2$ local gauge theory and thus the observables must be gauge invariant quantities. These quantities are the Wilson loops. In this theory, we make a Wilson loop $W^{x(z)}[\gamma]$ of the $x(z)$ type by drawing a closed string $\gamma$ on the lattice, and operating with $\sigma^x (\sigma^z)$ on all the spins encountered by the loop. In the polarized phase, the tension is high and it is difficult to create large loops. The expectation value of loops decays with the area enclosed by the loop. In the topologically ordered phase, large loops are less costly and their expectation value only decays at most with the perimeter of the loop. The phase transition is of the
confinement/deconfinement type. We can write any (contractible) Wilson loop as the product of some plaquette operator: $W^{x(z)}[\gamma]=\prod_{k\in S} B_k$In particular at the point $\tau=1$ when the model is the exact toric code, the expectation value of Wilson loops is $\langle|W^{x(z)}[\gamma]|\rangle =1$ for every loop $\gamma$, independently of its size. Of course, large loops are highly non-local observables.}
We have
computed the expectation value of {Wilson} loop operators of increasing size as a
function of $\tau $. As Fig.~\ref{tope1} shows, the expectation values
of large loops vanish in the spin-polarized phase, and increase
exponentially in the TO phase. However, in the limit of infinite
length, Wilson loops are not observables of 
the pure gauge theory \cite{measurability} and cannot be measured. 

{Nevertheless, topological order reveals itself in the way the
  ground state is entangled. If we compute the von Neumann entropy for
  a region with perimeter $L$, the entanglement entropy will be $S = L
  -1$ in the topological phase -- see Fig. \ref{tope3}. The spin
  polarized phase is not entangled. We see that there is a finite
  correction of $-1$ to the boundary law for the entanglement, which is due to the presence of topological order \cite{hiz1,Hamma:05a}}. Therefore we can consider as an alternative non local order parameter the
{\em topological entropy} \cite{Kitaev:06Levin:06}:
\begin{equation}
S_{\mathrm{top}}^{(R,r)}=S_{(A\cup B\cup C)}-S_{(A\cup C)}-S_{(A\cup
B)}+S_{A}
\end{equation}%
where $S_{\sigma}$ are the entanglement entropies associated with four
cuts $\sigma =\{A\cup B\cup C,\,A\cup C,A\cup B,A\}$, as depicted
in Fig.~\ref{lattice}.
We computed $S_{\mathrm{top}}(\tau )$ in the instantaneous ground state $|\psi
(\tau )\rangle $ for $L18$ and $L32$ ($L8$ is too small) in the ideal
model and for $L18$ in the perturbed model -- see
Figs.~\ref{tope2},\ref{tope3},\ref{tope4}.
In the spin-polarized phase, even for
finite systems,
$S_{\mathrm{top}}=0$ and it becomes different from zero only in the
vicinity of the critical point, after which it rapidly reaches
$1$ (as predicted in the thermodynamic limit in
Ref.~\cite{Kitaev:06Levin:06}).
To test whether $S_{\rm top}$ can discriminate between
symmetry breaking QPTs and TOQPTs, we show in in Fig.~\ref{tope2} the behavior of block
entanglement and $S_{\rm top}$ for a quantum {\em Ising model} in
$2D$. This model admits a QPT between a paramagnetic and
magnetically ordered phase, which is symmetry breaking. Block
entanglement detects the critical point sharply, while $S_{\rm top}$
does not (note the different scales on the left and right vertical
axes).
The small non-zero value of $S_{\rm top}$ is a finite size effect.

The block entropy in
Fig.~\ref{tope3} shows that the state is already rather
entangled in the spin-polarized region, whereas
$S_{\mathrm{top}}$ is almost zero before the transition to TO occurs.  Note that
the block entanglement at the critical point is bounded from above by
the final-state entanglement ($\tau =1$), which obeys the area
law. This is an example of the fact that in 2D, critical
systems do not need to violate the area law as in 1D. The useful
feature of $S_{\mathrm{top}}$ is not only that it can be used in order
to locate the critical point [Fig.~\ref{tope4}], but also that
it allows one to understand the type of QPT (symmetry breaking or TO).
Remarkably, Figs.~\ref{tope2},\ref{tope4} show that $S_{\mathrm{top}}$ has
these properties already for finite and very small
systems. The accuracy of the finite-size $S_{\mathrm{top}}$ at the
limit points $\tau=0,1$ is due to the fact that there the correlations are
exactly zero-ranged. This, however, is not the case for intermediate
$\tau$, especially near the QPT, so
how
$S_{\rm top}$ works as an order parameter, and how sharply its derivative
detects the QPT, are rather non-trivial.

In the presence of the perturbation $h^x(j)$, which tends to destroy
the loop structure, $S_{\mathrm{top}}$ detects the TOQPT up to the
value $P\sim 25$, after which a transition occurs: see Fig.~\ref{tope4}(inset). 
Overall, Figs.~\ref{tope3},\ref{tope4} show that the robustness of TO
against perturbations is a feature of the whole topological phase and
not only of the analytically solvable model at $\tau=1$.
Finally, we note another remarkable fact: setting the $x$-perturbation
$V$ to zero, and moving backward in time from $\tau=1$, we can view also the
tension term $H_{\xi }$ as a perturbation. This is due to the fact
that the toric code is symmetric under the exchange $x\leftrightarrow
z$ in the spin components.
The flatness of $S_{\mathrm{top}}$ in
Fig.~\ref{tope3} (squares and circles) shows the robustness of the
topological phase against this perturbation (see also
Ref. \cite{Trebst:06}).

\section{Conclusions}
We have presented a comprehensive numerical
study of a TOQPT. Our results show, using a variety of previously
proposed QPT detectors, that this is a second order transition.
Unlike the other detectors,
the topological entropy $S_{\rm top}$ is capable of distinguishing this TOPQT from a
standard one, already for small lattices. Strikingly, the model and its
TOQPT are highly robust against random perturbations not only deep
inside
the topological phase, where the gap protects
the ground state
from
perturbations, but
-- even more surprisingly --
at the gapless critical 
point. This phenomenon requires further investigation to be properly
understood. Moreover, $S_{\rm top}$  detects the TOQPT for perturbations
of strength up to 20\% 
of the strongest couplings. Of course finite-size effects can be
important, but it is not possible
at present
to compute
$S_{\rm top}$ exactly
without direct diagonalization, and this poses limits
on the
maximum size of
systems that can be studied.

We acknowledge financial support by DOE, Grant No.  DE-FG02-05ER46240
(to S.H.), and by NSF Grants No.  CCF-0523675 and CCF-0726439, and ARO-QA Grant
No. W911NF-05-1-0440 (to D.A.L). Computational facilities have been
generously provided by the HPCC-USC Center.

\end{document}